\documentstyle[11pt]{article}
\newcommand{\be}{\begin{equation}}
\newcommand{\ee}{\end{equation}}
\newcommand{\ba}{\begin{eqnarray}}
\newcommand{\ea}{\end{eqnarray}}

\begin{document}
\thispagestyle{empty}
\vspace{0.5in}
\begin{center}

{\Large \bf On nuclear matrix element uncertainties in short range 
$0\nu\beta\beta$ decay}

\vspace{1.0in}

{\bf H.V. Klapdor--Kleingrothaus and  H. P\"as}

\vspace{0.2in}

\noindent {\sl Max--Planck--Institut f\"ur Kernphysik\\
P.O. Box 103980, D--69029 Heidelberg,
Germany \\}

\vspace{1.0in}

\begin{abstract}
\noindent
The evaluation of short range contributions to neutrinoless double 
beta decay has been challenged due to critics of the ansatz of the nuclear 
matrix element calculations.
We comment on the critics and uncertainties of these calculations and the 
effect on the derived limits.
\end{abstract}
\end{center}

Neutrinoless 
double beta decay corresponds to the lepton-number converting
process
\be        
^{A}_{Z}X \rightarrow ^A_{Z+2}X + 2 e^-. 
\ee
So far no positive signal for this decay has been observed, 
yielding
the most stringent limit on the effective neutrino Majorana mass
and neutrino-mediated contributions from R-parity violating SUSY
and establishing this decay to be one of the most sensitive tools to search for
particle physics beyond the standard model. 
Besides these long-range contributions, where the decay is triggered by the 
exchange of a light Majorana neutrino
also contributions due to heavy particle exchange (superheavy 
neutrinos and SUSY partners) have been discussed, and extremely stringent
constraints on the effective superheavy neutrino mass $\langle m_H \rangle$
and R-parity violating coupling $\lambda'_{111}$
have been published (for an overview and recent limits see 
\cite{cosmo99,klapst}):
\ba
\langle m_H \rangle = \left|\sum_j \frac{U^2_{ej}}{m_j}\right|^{-1} 
> 9 \cdot 10^7 GeV\\
\lambda_{111}^{'}\leq 4\times 10^{-4}\Big(\frac{m_{\tilde{q}}}{100 
GeV} \Big)^2
\Big(\frac{m_{\tilde{g}}}{100 GeV} \Big)^{1/2}.
\ea 
For comparison, a future linear collider 
with a center of mass energy of
1 TeV would be sensitive to 250 TeV $< \langle m_H \rangle <$ 5000 TeV, only
\cite{minkneut} (for a serious discussion of possibilities to observe
inverse neutrinoless double beta decay at future colliders due to 
finetuned cancellations of mass eigenstates in the double beta decay 
observable see \cite{bela}).
These latter conclusions from the $0\nu\beta\beta$ decay 
half life limit
 have been challenged by critics \cite{minkcrit}
concerning the matrix 
element calculations  at short distances. 
In the following we will comment on the critics and uncertainties of 
these calculations and the effect on the derived limits.

The standard ansatz for nuclear matrix element calculations 
treats 
double beta decay in terms of nucleons of finite size with a hard core.
The finite nucleon size effect is taken into account by nucleon form factors
in momentum space \cite{45}
\be
F(q^2)=F(0)\left(1-\frac{q^2}{m_A^2} \right)^{-2}
\ee
with $m_A=0.85$ GeV. The form factors $F(0)$ used
have been calculated treating the quarks in the MIT bag model \cite{adl}.
The nucleon-nucleon repulsion at short distances is considered in two ways.
First, the repulsion effect is included in the nucleon potential.
In addition, to be conservative, the nucleon hard core is simulated 
with introducing a cutoff by multiplying the
two particle wave functions by the correlation function \cite{57}
\be
1-f(r)=1- e^{-ar^2}(1-b r^2).
\ee
The parameters $a$ and $b$ can be related to each other so that effectively,
there is one free parameter, the correlation length
\be
l_c= - \int_0^{\infty} ds ([1+f(r)]^2 -1). 
\ee 
The standard value of $l_c \simeq 0.7 $ fm fits experimental 
data from nucleon-nucleon scattering. In this approach
the total suppression of short range 
matrix elements compared to long range matrix elements 
with the same transition operator 
equals 1/20-1/30.

The dependence of short range nuclear matrix element calculations 
in the pn-QRPA model on the quantities 
$m_A$ and $l_C$ has been discussed  extensively in \cite{hirsch}
(for another recent calculation of the involved matrix elements, 
confirming the calculation in \cite{hirsch} with an accuracy of a factor of 2,
see \cite{jinr}). 
It has been shown that in this approach the main contribution to the matrix
element comes from nuclear distances larger than 1 fm. The matrix elements 
are stable to variations of $m_A$ and $l_C$, changes up to 50 \% of the 
standard values yield only comparable variations of the nuclear matrix 
elements. Although no guarantee - in the sense the nucleon can not be derived 
from QCD and no direct experimental test apart from comparison with data 
from nucleon-nucleon scattering is possible -
exists that this approach is applicable 
for the case of heavy particle exchange, it was successful in predicting 
the matrix element of the (long range) 
standard model mode of double beta decay 
(two neutrino emitting decay) with 
an accuracy of 
$\sqrt{2}$ (compare refs. \cite{50} and \cite{58}). 

The criticism of ref. \cite{minkcrit} is based on the argument that for 
intermediate particle masses 
as heavy as discussed here the correct picture would be the quark rather than 
the nucleon picture. One should keep in mind, however, that the heavy 
exchanged particles are virtual and that the momenta transferred are much 
smaller and that the quark dynamics are simulated by the effective treatment 
of nucleons with a form factor, hard core and nucleon-nucleon interaction. 
The total suppression 
of short range transitions compared to long range transitions
due to the quark-quark repulsion has been estimated in ref. \cite{minkcrit}
to yield a suppression by a factor of 1/40 or less. This estimation is based
on a spin singlet requirement to achieve an overall antisymmetric wave 
function ($\simeq 2/3$), the color Coulomb repulsion of the 
involved d-quarks  ($\simeq 1/3$ estimated by a WKB evaluation of the 
color Coulomb barrier) and a similar factor from the 
interaction of the remaining two quarks in the nucleus, which is justified by
the picture that each of the two decaying d-quarks is 
``pulled on by a u- and a d-quark from its own nucleon'', the latter being 
estimated
to be $\simeq (1/3)^2$ or less. Whether attracting
interactions between quarks belonging to the other nucleon change this picture 
is not discussed in ref. \cite{minkcrit}. Also effects of the nuclear 
environment may change this picture and are totally ignored in this 
estimation.
While this 
estimation is not based on an approach which is generally accepted (the
preprint ref. \cite{minkcrit} from 1996 is not published yet), the total 
suppression factor 1/40 argued, confirms the order of magnitude of the 
suppression
of short range matrix elements compared to long range matrix elements 
in the pn-QRPA approach
1/20-1/30. However, ref. \cite{minkcrit} incorrectly
applied this suppression factor 
to the limits derived with the pn-QRPA short range matrix elements and this way
considered the suppression factor two times. Moreover, old experimental 
limits have been used in the comparison of double beta decay and the 
inverse process. 

In fact the to our knowledge 
only serious attempt of a calculation based on a relativistic quark model,
see ref. \cite{faes},
confirms matrix element calculations in the standard approach with an accuracy 
of a factor of three.
It should be stressed also, that other decay modes, e.g. 
with pion exchange between 
the nucleons \cite{pion} and multiquark clusters in the nucleus
\cite{multi} have been considered yielding similar
results. 
We therefore assume it to be rather premature, to classify 
(as in \cite{minkneut}) all matrix elements calculated for heavy particle 
exchange as ``old'' in the sense of no more valid. 

If one in spite of these facts
assumes (incorrectly) 
the estimated suppression of short range matrix elements
from ref. \cite{minkneut},
the limit on the superheavy Majorana neutrino becomes 2000 TeV, still
being competitive to a 1 TeV linear collider. 
For supersymmetric contributions in addition one has to take into account that
the bound on the coupling scales with the square root of the nuclear matrix 
element, so that the estimated suppression would lead to a limit on 
$\lambda'_{111}$ 
being worse 
only by a factor of order 5. 

Summarizing, we commented on the critics of short range matrix element 
calculations for neutrinoless double beta decay. 
Since
a real alternative based on a treatment in the quark picture is missing
and in view of the lack 
of any reasonable estimation leading to considerably worse limits 
(i.e. more than a factor of three)  
we find it useful to 
present as limits furthermore the results of the calculations in the 
nucleon picture.
Moreover, even if one assumes the - clearly incorrect -
estimation of ref. \cite{minkcrit}, limits on SUSY are only worse by a 
factor of five and limits on superheavy neutrinos are still compatible
to what could be obtained at future linear colliders. It should be stressed
further, 
that these critics do not concern the neutrinoless double beta decay 
contributions with light particle exchange 
yielding limits on light neutrino masses \cite{numass},
R-parity violating SUSY \cite{rps}, 
leptoquarks \cite{lept} as well as violations of the equivalence 
principle and Lorentz invariance \cite{sark}.

\section*{Acknowledgement}
We thank M. Hirsch for useful discussions.

\end{document}